\documentclass[prd,superscriptaddress,altaffilletter,showpacs,nofootinbib]{revtex4}
\usepackage{amssymb}
\usepackage{amsmath}
\usepackage{graphicx}
\usepackage{subfigure}
\usepackage{epsfig}
\usepackage{epstopdf}
\usepackage[usenames,dvipsnames]{color}
\usepackage{dcolumn}
\usepackage{bm}

\begin{document}
\title{On localization of universal scalar fields in a tachyonic de Sitter thick braneworld}

\author{Alfonso D\'\i az-Furlong}\email{adiazfur@gmail.com}
\affiliation{Departamento de F\'{\i}sica, Universidad Aut\'onoma Metropolitana Iztapalapa,
San Rafael Atlixco 186, CP 09340, M\'exico D. F., M\'exico.}
\author{Alfredo Herrera-Aguilar}\email{alfredo.herrera.aguilar@gmail.com}
\affiliation{Departamento de F\'{\i}sica, Universidad Aut\'onoma Metropolitana Iztapalapa,
San Rafael Atlixco 186, CP 09340, M\'exico D. F., M\'exico.}
\affiliation{Instituto de F\'{\i}sica y Matem\'{a}ticas, Universidad Michoacana de San Nicol\'as de Hidalgo,\\
Edificio C-3, Ciudad Universitaria, CP 58040, Morelia, Michoac\'{a}n, M\'{e}xico.}
\author{Rom\'an Linares}\email{lirr@xanum.uam.mx}
\affiliation{Departamento de F\'{\i}sica, Universidad Aut\'onoma Metropolitana Iztapalapa,
San Rafael Atlixco 186, CP 09340, M\'exico D. F., M\'exico.}

\author{Refugio Rigel Mora-Luna}\email{rigel@fis.unam.mx}
\affiliation{Instituto de Ciencias F\'isicas, Universidad Nacional Aut\'onoma de M\'exico,\\
Apdo. Postal 48-3, 62251 Cuernavaca, Morelos, M\'exico.}

\author{Hugo A. Morales-T\'ecotl}\email{hugo@xanum.uam.mx}
\affiliation{Departamento de F\'{\i}sica, Universidad Aut\'onoma Metropolitana Iztapalapa,
San Rafael Atlixco 186, CP 09340, M\'exico D. F., M\'exico.}

\date{\today}

\begin{abstract}
Braneworld models may yield interesting effects ranging from high-energy physics to cosmology, or even some low-energy physics. Their mode structure modifies standard results in these physical realms that can be tested and used, for example,  to set bounds on the models parameters. Now, to define braneworld deviations from standard 4D physics, a notion of matter and gravity localization on the brane is crucial. In this work we investigate the localization of universal massive scalar fields in a de Sitter thick tachyonic braneworld generated by gravity coupled to a tachyonic bulk scalar field. This braneworld possesses a 4D de Sitter induced metric and is asymptotically flat despite the presence of a negative bulk cosmological constant, a novel and interesting peculiarity that contrasts with previously known models. It turns out that universal scalar fields can be localized in this expanding braneworld if their bulk mass obeys an upper bound, otherwise the scalar fields delocalize: The dynamics of the scalar field is governed by a Schr\"odinger equation with an analog quantum mechanical potential of modified P\"{o}schl-Teller type. This potential depends on the bulk curvature of the braneworld system as well as on the value of the bulk scalar field mass. For masses satisfying a certain upper bound, the potential displays a negative minimum and possesses a single massless bound state separated from the Kaluza-Klein (KK) massive modes by a mass gap defined by the Hubble (expansion scale) parameter of the 3-brane. As the bulk scalar field mass increases, the minimum of the quantum mechanical potential approaches a null value and, when the bulk mass reaches certain upper bound, it becomes positive (eventually transforming into a potential barrier), leading to delocalization of the bulk scalar field from the brane. We present analytical expressions for the general solution of the Schr\"odinger equation. Thus, the KK massive modes are given in terms of general Heun functions as well as the expression for the massless zero mode, giving rise to a new application of these special functions.
\end{abstract}





\maketitle

\section{Introduction and review of the model}

The study of braneworlds has considerably evolved in the last years.
Within the braneworld paradigm theoretical science has found
consistent alternatives to solve or reformulate various problems of
modern physics, such as the cosmological constant problem 
\cite{RubakovPLB1983136}--\cite{Randjbar-DaemiPLB1986}, 
the gauge hierarchy problem \cite{AntoniadisPLB1990}--\cite{Lykken}
or possible low energy effects including the Lamb shift in Hydrogen 
\cite{Lamb}, the Casimir force \cite{Casimir} and non-singular particle 
potentials \cite{non-singular}, among others (for an interesting review 
see \cite{1004.3962}). In this context, there are many models of thin
branes \cite{rs,gog} and thick branes \cite{0904.1775} which shape
the fifth dimension using different matter fields living in the
bulk. The freedom we have to use matter fields in the bulk has led
some authors to use all kinds of bulk fields, scalar fields being
the most popular among them \cite{1004.3962}--\cite{germanetal}.

In order to model the fifth dimension, the use of a 5D tachyonic
scalar field was explored in \cite{NonLocalizedFermion2,germanetal},
however along this research line, some attempts failed to localize
4D gravity (and other matter fields) or to find real solutions to
the system of highly nonlinear equations. In fact, the tachyon
condensate proposed by A. Sen \cite{sen} presents an additional
difficulty due to the complex form of the field equations, however,
despite these difficulties, in a recent work \cite{germanetal}, an
interesting and novel solution which localizes 4D gravity using a
real tachyon field was obtained. Subsequently, the localization of
gauge vector fields in this braneworld was recently accomplished and
reported in the literature \cite{1401.0999} (see \cite{Corradini} as well for
an alternative mechanism for localizing gauge and Kalb--Ramond fields
on thick branes based on a 5D Stueckelberg action, and \cite{Proca}--\cite{FJ} 
for its canonical quantization). A peculiar feature of
this expanding thick braneworld model is that both the gravitational
and gauge vector fields only possess a {\it single} massless bound
state localized on the 3-brane in their corresponding mass spectra,
allowing us to successfully recover the phenomenology of our 4D
world. This expanding de Sitter braneworld is also interesting from
the cosmological point of view since it qualitatively describes both
the inflationary period of the early Universe and the accelerating
expansion that we observe nowadays. In particular, since the cosmic
inflation theory is in quite good agreement with the properties of
the temperature fluctuations observed in the Cosmic Microwave
Background Radiation, and since inflation likely took place at very
high temperatures and, hence, its study involves some assumptions
related to the relevant physics that takes place at such high
energies, cosmologists have performed several attempts to construct
consistent inflationary models within string theory and supergravity
\cite{popeetal,burgess,SC}.

In this work we will continue with the study of localization of
universal massive scalar fields (for localization of matter see
\cite{BajcPLB2000}--\cite{Guo_jhep} ), on the aforementioned
tachyonic thick braneworld \cite{germanetal}, following
\cite{1401.0999}, where the possibility of localizing bulk gauge
vector fields was successfully explored. This paper is organized in
three parts. In the first section we present an introduction and a
brief review of the derivation of the de Sitter tachyonic thick
braneworld, together with some of its relevant physical properties.
In the second section we study the localization of massive universal
scalar fields, and we find that the massless zero mode is localized
on the brane only in the case when its 5D mass obeys an upper bound,
otherwise delocalizing from it; on the other hand, in the mass
spectrum there is also a tower of Kaluza--Klein (KK) delocalized
massive modes separated from the massless one by a mass gap. The
corresponding expressions for both the massless and KK massive modes
are given in terms of general Heun functions. Finally, in the third
section we present our conclusions and discuss a little bit on the
obtained results.

\subsection{Review of the tachyonic de Sitter thick braneworld }
\label{SecModel}

Let us start with the following 5D action for a thick braneworld model that emerges from the interaction of gravity with a bulk tachyonic scalar field in the presence of a bulk cosmological constant \cite{germanetal}
\begin{equation}
S = \int d^5 x \sqrt{-g} \left(\frac{1}{2\kappa_5^2} R -
\Lambda_5 - V(T)\sqrt{1+g^{AB}\partial_{A} T\partial_{B} T} \right),
\label{action}
\end{equation}
where $R$ is the 5D scalar curvature, $\Lambda_{5}$ is the bulk
cosmological constant, and $\kappa_5^2=8\pi G_5$ with $G_5$ the
five--dimensional Newton constant. Here we use the signature
$(-++++)$ and the Riemann tensor, defined as follows
$R_{MNPQ}=\frac{\Lambda_5}{6}\left(g_{MP}g_{NQ}-g_{MQ}g_{NP}\right)$,
gives rise to the Ricci one $R_{NQ}=R^M_{NMQ}$ upon contraction of
its first and third indices, where $M,N,P,Q=0,1,2,3,5.$ From this
action, the Einstein equations with a cosmological constant in five
dimensions are
\begin{eqnarray}\label{EinsteinEq_5d}
G_{AB} = - \kappa_5^2 ~\Lambda_5 g_{AB} + \kappa_5^2
~T_{AB}^{\it{bulk}}
\end{eqnarray}
where $T_{AB}^{\it{bulk}}$ is the energy--momentum tensor for the bulk tachyonic scalar field.

A 5D metric ansatz compatible with an induced 3--brane with spatially flat cosmological background can be taken to be
\begin{eqnarray}\label{metricw}
 ds^2 = e^{2f(w)} \left[- d t^2 + a^2(t) \left(d x^2 + d y^2 + d
z^2 \right)+ d w^2 \right],
\end{eqnarray}
where $\text{e}^{2f(w)}$ and $a(t)$ are the warp factor and the scale factor of the brane and $w$ stands for the extra dimensional coordinate. The matter field equation is obtained by variation of the action with respect to the tachyonic scalar field and it is expressed in the following form:
\begin{equation}
\partial_{A}\left[\frac{\sqrt{-g} V(T) \partial^A T}{\sqrt{1+  (\nabla T)^2}}
 \right ]
- \sqrt{-g} \sqrt{1+  (\nabla T)^2} \frac{\partial V(T)}{\partial T} = 0.
\end{equation}
Thus, the braneworld field system consists of the bulk tackyonic scalar field, its self--interaction potential, and the metric functions which parameterize the line element (\ref{metricw}).

The full solution for the metric scale and warp factors respectively reads
\begin{equation}
a(t)=e^{H\,t}, \qquad  \qquad
f(w)=\frac{1}{2}\ln\left[s\,{\rm sech}\left(\,H\,(2w+c)\right)\right],
\label{scalewarpfactors}
\end{equation}
whereas the tachyon scalar field is given by
\begin{eqnarray}
T(w) &=& \pm\sqrt{\frac{-3}{2\,\kappa_5^2\,\Lambda_5}}\
\mbox{arctanh}\left(\frac{\sinh\left[\frac{H\,\left(2w+c\right)}{2}\right]}
{\sqrt{\cosh\left[\,H\,(2w+c)\right]}}\right)  \nonumber \\
&=& \pm\sqrt{\frac{-3}{2\,\kappa_5^2\,\Lambda_5}}\ \ln\left[
\mbox{tanh}\left(\frac{H\,\left(2w+c\right)}{2}\right)+\sqrt{1+\mbox{tanh}^2\left(\frac{H\,\left(2w+c\right)}{2}\right)}
\right] , \label{Tw}
\end{eqnarray}
while  the tachyon potential has the form
\begin{eqnarray}
V(T) &=& - \Lambda_5\
\mbox{sech}\left(\sqrt{-\frac{2}{3}\kappa_5^2\,\Lambda_5}\ T\right)
\sqrt{6\ \mbox{sech}^2\left(\sqrt{-\frac{2}{3}\kappa_5^2\,\Lambda_5}\ T\right)-1}\nonumber \\
&=& - \Lambda_5\ \sqrt{\left(1 + \mbox{sech}\left[\,H\,(2w+c)\right]\right)
\left(1+\frac{3}{2}\ \mbox{sech}\left[\,H\,(2w+c)\right]\right)},
\label{VT}
\end{eqnarray}
where $H$, $c$ and $s>0$ are arbitrary constants. We should mention that we have set
\begin{equation}
s=-\frac{6H^2}{\kappa_5^2\,\Lambda_5}, \label{s}
\end{equation}
with a negative bulk cosmological constant $\Lambda_5<0$ when deriving the last two equations.

The corresponding curvature scalar reads
\begin{equation}
R=-\frac{14}{3}\kappa_5^2\,\Lambda_5\,\mbox{sech}\left[\,H\,(2w+c)\right].
\label{R}
\end{equation}
As one can easily see, this invariant is positive definite (since the bulk cosmological constant is negative) and asymptotically flat along the fifth dimension. Thus, the thick braneworld under consideration interpolates between two 5D Minkowski spacetimes and is completely free of naked singularities that usually arise when the mass spectrum of Kaluza--Klein tensorial (graviton) fluctuations displays a mass gap, an appealing feature from the phenomenological viewpoint, as in this model (see \cite{0910.0363} for details).

\section{Localization of universal scalar fields on the tachyonic de Sitter brane}
\label{SecLocalization}

In this section we shall study the localization of bulk massive scalar fields on the above presented tachyonic de Sitter thick braneworld through gravitational interaction. It should be mentioned that we have implicitly assumed that the universal scalar fields considered below do make a small contribution to the energy of the bulk in such a way that the aforementioned solution to the field system remains valid in the presence of bulk scalar matter. Thus, the massive scalar fields we shall consider possess a quite small mass in order for this approximation to be valid. We shall further discuss the structure of the mass spectrum of the massive scalar field by analyzing the analog quantum mechanical potential of the corresponding Schr\"{o}dinger equation for their KK modes and we shall present their general solution in terms of general Heun functions.

Thus, let us consider the action of a massive real scalar field minimally coupled to gravity
\begin{eqnarray}\label{action_scalar}
S=\int d^{5}x\sqrt{-g}~
          \left(-\frac{1}{2}\,g^{MN}\partial_{M}\Phi\partial_{N}\Phi-\frac{m^2_{s}}{2}\Phi^2\right),
\end{eqnarray}
where $m_s$ is the bulk scalar field mass. With the aid of the metric (\ref{metricw}), the equation of motion derived
from (\ref{action_scalar}) reads
\begin{eqnarray}\label{EqOfScalar5D}
\frac{1}{\sqrt{-\hat{g}}}\partial_{\mu}\left(\sqrt{-\hat{g}} \hat{g}^{\mu \nu}\partial_{\nu}
\Phi\right) + e^{-3f}\partial_{w} \left(e^{3f}\partial_w\Phi\right)  = 0.
\end{eqnarray}
We further propose the following ansatz for theKK decomposition of the scalar field $\Phi(x,w)=\sum_{n}\phi_{n}(x)\chi_{n}(w)e^{-3f/2}$ and demand that its 4D profile $\phi_{n}$ obeys a massive Klein--Gordon equation:
\begin{eqnarray}
\label{4dKGEq}
\left[\frac{1}{\sqrt{-\hat{g}}}\partial_{\mu}\left(\sqrt{-\hat{g}}
   \hat{g}^{\mu \nu}\partial_{\nu}\right) -m_{n}^{2} \right]\phi_{n}(x)=0,
\end{eqnarray}
where $m_{n}$ is the 4D mass of the scalar field. On the other hand, the scalar KK mode $\chi_{n}(z)$ must satisfy a Schr\"{o}dinger equation:
\begin{eqnarray}
\left[-\partial^{2}_w+ V_{s}(w)\right]{\chi}_n(w)
  =m_{n}^{2} {\chi}_{n}(w),
  \label{SchEqScalar1}
\end{eqnarray}
with the following effective quantum mechanical potential
\begin{eqnarray}
  V_s(w)=\frac{3}{2} f'' + \frac{9}{4}f'^{2} + e^{2f}m_s^2.
  \label{VScalar}
\end{eqnarray}
It is evident that this potential only depends on bulk entities: the warp factor and the mass of the universal scalar field.

Thus, under the considered separation of variables, the 5D action (\ref{action_scalar}) yields a standard 4D action for a massless and a series of massive scalar modes
\begin{eqnarray}
 S=- \frac{1}{2} \sum_{n}\int d^{4} x \sqrt{-\hat{g}}
     \bigg(\hat{g}^{\mu\nu}\partial_{\mu}\phi_{n}
           \partial_{\nu}\phi_{n}
           +m_{n}^2 \phi^{2}_{n}
     \bigg), \label{ScalarEffectiveAction}
\end{eqnarray}
after integrating along the fifth dimension, and requiring the fulfillment of both the following orthonormalization conditions
\begin{eqnarray}
 \int^{\infty}_{-\infty}
 \;\chi_m(w)\chi_n(w) dw = \delta_{mn}
 \label{normalizationCondition1}
\end{eqnarray}
and the Schr\"{o}dinger equation (\ref{SchEqScalar1}).

Finally, for the warp factor solution (\ref{scalewarpfactors}) the analog quantum mechanical potential adopts the form
\begin{equation}
V_s=\frac{3H^{2}}{4}\left[3-7{\rm
sech}^{2}(2Hw)\right] + s\,m^2_{s}\,{\rm sech}(2Hw). \label{V0}
\end{equation}

\vskip 3mm

\begin{figure}[htb]
\begin{center}
\includegraphics[width=7cm]{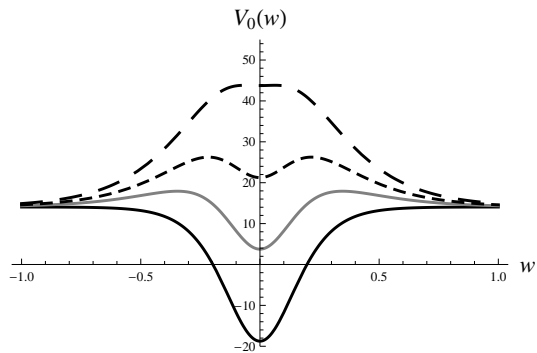}
\end{center}\vskip -5mm
\caption{The profile of the  potential whit $m_{s}=0$ (thick line),
$m_{s}=3$ (thin line), $m_{s}=4$ (dashed small lines), $m_{s}=4$
(dashed large lines), along the fifth dimension. Here we have set
$s=2.5$, $H=2.5$ and $k_1=1$ for simplicity.} \label{V0mass}
\end{figure}
\begin{figure}[htb]
\begin{center}
\includegraphics[width=7cm]{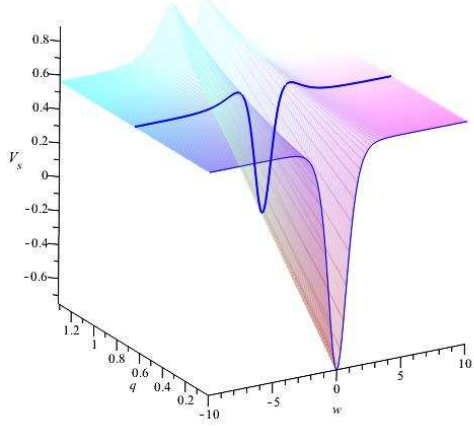}
\end{center}\vskip -5mm
\caption{The 3D surface shows the continuous shape of the potential
for each point $(x,q)$ with $x=2Hw$,
 $q=\frac{3m^2_{s}}{2k^{2}_{5}\Lambda_{5}}$ where the zero mode is
located in the 2D domain surface
$(-\infty,\infty)\times[0,\frac{21}{32})$, the first position where
 the zero KK mode is located in the potential is given by the thin blue line
bounded by $m_ {s}=0$, and the least position where
 the zero  KK
mode is located in the potential is given by the thick blue line
bounded by the critical value $m^{2}_{s}=\frac{7k^{2}_5
\Lambda_{5}}{8}$.}\label{pot3d}
\end{figure}
In general, this potential can have a minimum or  maximum at $w=0$,
and its behavior is governed by the dominant second or third term on
the right hand side of (\ref{V0}) as depicted in Fig. \ref{V0mass}.
Generally, when $0\le m^2_{s}<-\frac{7}{8}k^2_{5}\Lambda_{5}$ the
dominant term is the second one and the potential is of modified
P\"{o}schl-Teller type with a massless KK state localized on the
brane. However, as $m^2_s$ increases, the minimum of the
P\"{o}schl-Teller potential approaches a zero value and then becomes
positive (namely, when $m^2_{s}\geq-\frac{7}{8}k^2_{5}\Lambda_{5}$),
leading to a P\"{o}schl-Teller potential with no bound states which
eventually becomes a potential barrier, making evident that there
are not localized modes at all.

Thus, it is clear that in this case, the mechanism of universal
scalar field localization depends not only on the geometry of bulk,
but also on the bulk mass of the scalar field $m_{s}$, a parameter
that plays a crucial role in the sense that it separates two
localization phases: when it is small enough, leads to the
localization of the massless zero mode on the brane,
notwithstanding, when it increases and reaches a critical value, it
delocalizes the full mass spectrum pushing all (massless and
massive) KK modes outside the brane. Particularly, when
$m_{n}=m_{s}=0$ the potential well reaches its maximum depth and has
a single localized bound state. We also can see that keeping the
mass $m_{n}=0$ and increasing the value of the mass in the bulk, the
zero mode is slowly delocalized from the brane as $m_{s}$ approaches
the value $m^2_{s}=-\frac{7}{8}k^2_{5}\Lambda_{5}$ (see Fig.
\ref{pot3d}).

 Before presenting the
solution corresponding to the KK mass spectrum problem on the
equation (\ref{SchEqScalar1}) we will make the following change of
variable $x=2Hw$  then we shall the following anzats for the
function $\chi(x)_{n}={\rm sech^{\frac{3}{4}}}(x)y(x)_{n}$, finally
redefining the variable $ x={\rm arcsech(z)}$ the equation
(\ref{SchEqScalar1}) takes the following form
\begin{equation}
\frac{d^{2}y(z)_{n}}{dz^2}+\frac{\left(7z^2-5\right)}{2z(z-1)(z+1)}
\frac{dy(z)_{n}}{dz}+\frac{\left(\frac{3m^{2}_{s}}{2k^{2}_{5}\Lambda_{5}}z-\frac{m^{2}_{n}}{4H^2}\right)}{z^2(z-1)(z+1)}y(z)_{n}=0\label{dfeqn}
\end{equation}

 The general solution to the equation (\ref{dfeqn}) with the
 reads
\begin{eqnarray}\label{heunsol} \nonumber
y(z)_{n}&=&  \left[
K_{1}\,z^{\frac{-3H+\sqrt{9H^2-4m^{2}_{n}}}{4H}}(1-z)^{\frac{1}{2}}{\rm HeunG}\left(-1,a_{+},b_{+},c_{+},d_{+},\frac{1}{2},-z\right)\right. \\
&+& \left.
K_{2}\,z^{-\frac{3H+\sqrt{9H^2-4m^{2}_{n}}}{4H}}(1-z)^{\frac{11}{4}}{\rm
HeunG}\left(-1,a_{-},b_{-},c_{-},d_{-},\frac{1}{2},-z\right))\right],
\end{eqnarray}
where $K_{1}$ and $K_{2}$ are arbitrary constants, HeunG are general
Heun special functions and the coefficients of the solution
(\ref{heunsol}) are defined by
\begin{equation}\nonumber
a_{\pm}=\frac{2Hk^{2}_{5}\Lambda_{5}\pm
\sqrt{9H^2-4m^2_{n}}}{4Hk^{2}_{5}\Lambda_{5}}+q, \qquad
 b_{\pm}=\frac{9H\pm\sqrt{9H^2-4m^2_{n}}}{4H}, \qquad
 c_{\pm}=-\frac{H\pm\sqrt{9H^2-4m^2_{n}}}{4H},
\end{equation}
\begin{equation}\nonumber
 d_{\pm}=\frac{2H\pm\sqrt{9H^2-4m^2_{n}}}{2H}.
\end{equation}

Then the whole solution has the form $\chi_{n}={\rm
sech^{\frac{3}{4}}(x)y(x)_{n}}$.

\begin{figure}[htb]
\begin{center}
\includegraphics[width=7cm]{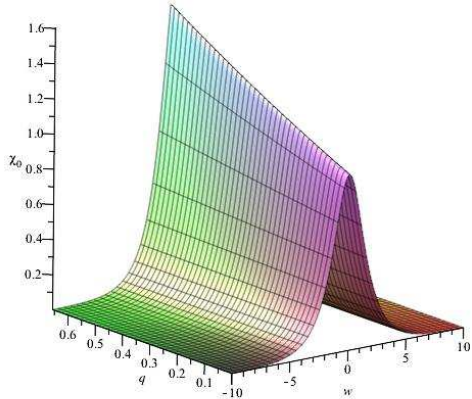}
\end{center}\vskip -5mm
\caption{This 3D figure shows the  shape for the zero mode in the
continuous range for $0\leq q<\frac{21}{32}$ which corresponds to
the critical values $m_{s}=0$ and
$m^{2}_{s}=\frac{7k^{2}_{5}\Lambda_{5}}{8}$.}\label{masslessmode}
\end{figure}

 In particular, for the massless
zero mode bound state (with $m_n=0$) starting from (\ref{dfeqn}) we
have the general Heun equation
\begin{equation}
\frac{d^{2}y(z)_{0}}{dz^2}+\frac{\left(7z^2-5\right)}{2z(z-1)(z+1)}
\frac{dy(z)_{0}}{dz}+\frac{q}{z(z-1)(z+1)}y(z)_{0}=0,\label{Heung}
\end{equation}

where $q=\frac{3m^{2}_{s}}{2k^{2}_{5}\Lambda_{5}}$. Now we can
recast the zero mode solution solving the general Heung equation
(\ref{Heung})
\begin{eqnarray}\label{yheunsol0} \nonumber
y(z)_{0}&=&\left[
C_{1}\,z^{-\frac{3}{2}}{\rm HeunG}\left(-1,q,-\frac{3}{2},1,-\frac{1}{2},\frac{1}{2},-z\right)\right. \\
&+& \left. C_{2}\,{\rm
HeunG}\left(-1,q,0,\frac{5}{2},\frac{5}{2},\frac{1}{2},-z\right)\right],
\end{eqnarray}
the complete solution for the zero mode in the $x$ coordinate reads
\begin{eqnarray}\label{chiheunsol0} \nonumber
\chi_{0} &=& {\rm sech}^{\frac{3}{4}}(x) \left[
C_{1}\,{\rm cosh^{\frac{3}{2}}(x)}{\rm HeunG}\left(-1,q,-\frac{3}{2},1,-\frac{1}{2},\frac{1}{2},-{\rm sech}(x)\right)\right. \\
&+& \left. C_{2}\,{\rm
HeunG}\left(-1,q,0,\frac{5}{2},\frac{5}{2},\frac{1}{2},-{\rm
sech}(x)\right)\right].
\end{eqnarray}
 However, only
the second term corresponds to a localized configuration, therefore
we need to set $C_1=0$. The corresponding extra dimensional profile
of this zero mode is displayed in Fig. \ref{masslessmode} within the
range of allowed values for the bulk mass of the scalar field, i.e.
for $0\le m^2_{s}<-\frac{7}{8}k^2_{5}\Lambda_{5}$, which in turn
implies the following bounds for the $q$ parameter $0\le q < 21/32$.

Within the framework of a little bit less general case we can consider that $m_{s}\ll1$ in order to guarantee the localization of the only discrete state of the KK mass spectrum on the brane. Thus, by neglecting the small parameter $m_{s}$, our potential (\ref{V0}) adopts the approximate form
 \begin{equation}
V_s=\frac{3H^{2}}{4}\left[3-7{\rm sech}^{2}(2Hw)\right].
\label{V0sms}
\end{equation}

\begin{figure}[htb]
\begin{center}
\includegraphics[width=7cm]{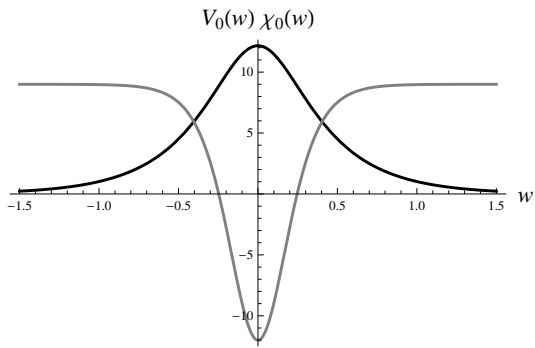}
\end{center}\vskip -5mm
\caption{The profile of the modified P\"oschl--Teller potential
(thin line) and the localized  zero mode (thick line) along the
fifth dimension. Here we have set  $H=2$ and $\alpha_{1}=10$ for
simplicity.} \label{VqmO}
\end{figure}

This approximated potential still has  a mass gap given by $\frac{9 H^{2}}{4}$ in the spectrum of KK massive fluctuations of the scalar field. By redefining the extra coordinate $u=2Hw$ the Schr\"{o}dinger equation (\ref{SchEqScalar1}) transforms into  \cite{PRD0709.3552}
\begin{equation}
 \left(-\partial^{2}_{u} - \frac{21}{16}{\rm sech}^{2}u \right)\chi(u) =
 \left(\frac{m^{2}_n}{4H^{2}} - \frac{9}{16}\right)\chi(u),
 \label{Scheqnu}
\end{equation}
and adopts the canonical form of the classical eigenvalue problem for the Schr\"odinger equation with a quantum mechanical potential of modified P\"oschl--Teller type:
\begin{equation}
 \left[-\partial^{2}_{u} - n(n+1){\rm sech}^{2}u \right]\chi(u) =
 E\,\chi(u)
 \label{Schcan}
\end{equation}
where $n=3/4$ and now $E=\frac{m^{2}_n}{4H^{2}} - \frac{9}{16}$. Since $n<1$, then $[n]=0$ and, as it is already known, there is just one bound state, the massless one, in the mass spectrum of the KK fluctuations of the universal scalar field. It turns out that (\ref{Scheqnu}) can be integrated for any arbitrary mass and possess the following general solution:
\begin{equation}
 \chi(w)=c_1\,P^{\mu}_{3/4}\left(\tanh(2Hw)\right)+
 c_2\,Q^{\mu}_{3/4}\left(\tanh(2Hw)\right)
\end{equation}
where $c_1, c_2$ are integration constants, and $P^{\mu}_{3/4}$ and $Q^{\mu}_{3/4}$ are associated Legendre functions of first and second kind, respectively, with degree $\nu=3/4$ and order $\mu=\sqrt{\frac{9}{16}-\frac{m^{2}_n}{4H^{2}}}$. It turns out that for the
massless zero mode $\mu=\nu=3/4$ and the exact solution can be written as follows
\begin{equation}
 \chi_{0}(w)=-\alpha_1\,\left[P^{3/4}_{3/4}\left(\tanh(2Hw)\right)+
 \frac{2}{\pi}\,Q^{3/4}_{3/4}\left(\tanh(2Hw)\right)\right],
 \label{zeromode}
\end{equation}
where $c_1=-\alpha_{1}$, $\alpha_{1} > 0$ and we have chosen
$c_2=2c_1/\pi$ in order to obtain a localized field configuration
along the fifth dimension. This massless bound state corresponds to
the localized zero mode of the scalar field on the 3--brane. The
profile of both the modified P\"oschl--Teller potential and the
massless zero mode of the scalar field along the extra dimension is
displayed in Fig. \ref{VqmO}.

\section{Conclusions and discussion}
\label{SecConclusion}

Extra dimensional models like braneworlds are of interest to look for new physics not only because of the possible deviations from standard physics they may lead to, but because they may inspire further, perhaps more precise tests, not previously noticed; this was the case, for instance, for the experiments proving Newton's law at the submillimeter length scale. In order to elucidate any new effect, ones has to contrast it with 4D data and/or observations and tipically this emerges within the model through a localization mechanism for matter and gravity. 

In this work we have investigated the localization properties of massive universal scalar field in a tachyonic de Sitter thick braneworld which interpolates between two Minkowski 5D spacetimes. The dynamics of the scalar field is governed by a Schr\"{o}dinger equation with an analog quantum mechanical potential completely determined by the 5D warping and the 5D mass of the free scalar field. We have found that this bulk mass plays the role of  a continuous parameter which can delocalize all the mass spectrum of the scalar field, i.e. the massless bound state and the massive KK ones, from the expanding 3-brane when it reaches a critical value. Thus, we have uncovered two localization phases of the 4D massless zero mode depending on how small is the mass of the bulk scalar field compared to a product of the 5D cosmological and gravitational constants. In fact, a result of this nature is to be expected due to the assumption we made that the 5D scalar field does not interact with the gravitational field in the bulk leaving the background solution valid, and neglecting its gravitational back-reaction.

On the basis of the performed analysis of the structure of the analog quantum mechanical potential one can easily see how sensitive is our model, where the coupling constant $m_{s}$ can  drag  the 4D massless zero mode of the scalar  field outside the  brane. Thus, the matter density on the bulk must be located far away from the brane in order to make a small contribution to the parameter $m_{s}$ in such a way that we could have the KK massless zero  mode localized inside the brane. It is worth noticing that, usually, when one one takes into account the gravitational back-reaction, the scalar modes (both massless and massive) are not localized on the brane anymore, in accordance with the previously described result since the 5D mass of the scalar field can be interpreted as a measure of its gravitational back-reaction.

A similar and interesting situation arises within the context of the AdS/CFT correspondence: when computing the amplitudes of a scalar field, the effect of its interaction with its dual operator depends on the quotient $\Big({energy\over mass}\Big)^{\alpha}$, for some constant $\alpha>0$. Thus, three different situations arise, depending on the value of the scalar field mass $m^2$, which in an AdS space is allowed to acquire negative values as far as the so-called Breitenlohner-Freedman bound is satisfied.
It turns out that for $m^2>0$ the effects of the interaction of the scalar field with its dual operator, called irrelevant in this case, can be neglected for low energies since they are suppressed by powers. Therefore, the interaction goes away in the IR, but changes completely the UV sector of the theory.
When $m=0$ the corresponding dual operator is called marginal. Finally, if  $m^2$ adopts the allowed negative values, the dual operator is called relevant and changes the IR sector of the theory (see, for instance, \cite{Ramallo} for a clear exposition of this issue).

Thus, along with the gravitational and gauge vector fields, the 4D scalar fields can be localized on the considered tachyonic de Sitter thick braneworld, if the mass of the bulk scalar field is bounded from above. It remains to determine whether fermions can be localized on this kind of braneworld configurations, however, one should first overcome the technical difficulties that this models presents in such a study. In the case of fermion localization on the tachyonic expanding 3-brane, one would be able to compute the corrections to Coulomb law coming from the KK massive modes of the gauge vector field in the non-relativistic limit. Since the photon mass is very suppressed, this study could lead, in principle, to establish whether this braneworld model is viable or not from the phenomenological point of view.

\section*{Acknowledgements}

AHA and RRML acknowledge useful and illuminating discussions with Dagoberto Malag\'on Morej\'on. AHA is grateful to ICF, UNAM and UAM-I for hospitality, as well as to \textquotedblleft Programa de Apoyo a Proyectos de Investigaci\'on e Innovaci\'on Tecnol\'ogica\textquotedblright\, (PAPIIT) UNAM, IN103413-3, {\it Teor\'ias de Kaluza-Klein, inflaci\'on y perturbaciones gravitacionales} for financial support. ADF gratefully acknowledge support from the PROMEP
program {\it Becas Posdoctorales en Cuerpos Acad\'emicos consolidados y en consolidaci\'on} under the grant No. 12312096. RRML acknowledges a postdoctoral grant from CONACyT at ICF-UNAM. All authors thank SNI.


\begin{thebibliography}{99}



\bibitem{RubakovPLB1983136}
 V.A. Rubakov and M.E. Shaposhnikov,
    {\it Do we live inside a domain wall?},
    Phys. Lett.  B {\bf 125} (1983) 136;
 V.A. Rubakov and M.E. Shaposhnikov,
    {\it Extra space-time dimensions: towards a solution to
    {     the cosmological constant problem}},
    Phys. Lett.  B {\bf 125} (1983) 139.

\bibitem{Randjbar-DaemiPLB1986}
 S. Randjbar-Daemi and C. Wetterich,
   {\it Kaluza-Klein solutions with noncompact internal spaces},
   Phys. Lett.  B {\bf 166} (1986) 65.

\bibitem{AntoniadisPLB1990}
 I. Antoniadis,
    {\it A possible new dimension at a few Tev},
    Phys. Lett. B {\bf 246} (1990) 377.

\bibitem{ADD}
 N. Arkani-Hamed, S. Dimopoulos and G. Dvali,
    {\it The hierarchy problem and new dimensions at a millimeter},
    Phys. Lett.  B {\bf 429} (1998) 263;
 I. Antoniadis, N. Arkani-Hamed, S. Dimopoulos and G. Dvali,
    {\it New dimensions at a millimeter to a Fermi and superstrings at a TeV},
    Phys. Lett.  B {\bf 436} (1998) 257.
    
\bibitem{gog}
    M. Gogberashvili,
               {\it Hierarchy problem in the shell universe model},
               Int. J. Mod. Phys. D {\bf 11} (2002) 1635,
               {\it Four dimensionality in noncompact Kaluza-Klein
               model},
               Mod. Phys. Lett. A {\bf 14} (1999) 2025.

\bibitem{rs}
 L. Randall and R. Sundrum,
    {\it A Large Mass Hierarchy from a Small Extra Dimension},
    Phys. Rev. Lett. {\bf83} (1999) 3370;
    {\it An alternative to compactification},
    Phys. Rev. Lett. {\bf83} (1999) 4690.

\bibitem{Lykken}
 J. Lykken and L. Randall,
    {\it The Shape of Gravity},
    JHEP {\bf 0006} (2000) 014.

\bibitem{Lamb}
   H.A. Morales-T\'ecotl, O. Pedraza and L.O. Pimentel,
       {\it Low-energy effects in brane worlds: Liennard-Wiechert potentials and Hydrogen Lamb shift},
       Gen. Rel. Grav. {\bf 39} (2007) 1185. 

\bibitem{Casimir}
	R. Linares, H.A. Morales-T\'ecotl and O. Pedraza,
	{\it Casimir force in brane worlds: coinciding results from GreenÕs function and Zeta function approaches},
	Phys. Rev. D {\bf 81} (2010) 126013. 

\bibitem{non-singular}
R. Linares, H.A. Morales-T\'ecotl and O. Pedraza, 
{\it Gravitational potential of a point mass in a brane world} 
Phys. Rev. D {\bf 86} (2014) 066002.

\bibitem{1004.3962}
 R. Maartens and K. Koyama,
        {\it Brane--World Gravity}, Living Rev. Rel. {\bf13}
        (2010) 5.

\bibitem{0904.1775}
 V. Dzhunushaliev, V. Folomeev, and M. Minamitsuji,
    {\it Thick brane solutions},
    Rept. Prog. Phys. {\bf 73} (2010) 066901.

\bibitem{De_Wolfe_PRD_2000}
 O. DeWolfe, D.Z. Freedman, S.S. Gubser and A. Karch,
    {\it Modeling the fifth dimension with scalars and gravity},
    Phys. Rev. D {\bf 62} (2000) 046008.

\bibitem{Gremm_2000}
 M. Gremm,
    {\it Four-dimensional gravity on a thick domain wall},
    Phys. Lett. B {\bf 478} (2000) 434;

\bibitem{stabilitythbws}
 K. Ghoroku and M. Yahiro,
    {\it Instability of thick brane worlds},
    hep-th/0305150;
 S. Kobayashi, K. Koyama and J. Soda,
    {\it Thick brane worlds and their stability},
    Phys. Rev. D {\bf 65} (2002) 064014.

\bibitem{KobayashiPRD2002}
 S. Kobayashi, K. Koyama and J. Soda,
    {\it Thick brane worlds and their stability},
    Phys. Rev. D {\bf 65} (2002) 064014.

\bibitem{Csaki_NPB_2000}
 C. Csaki, J. Erlich, T. Hollowood and Y. Shirman,
    {\it Universal Aspects of gravity localized on thick branes},
    Nucl. Phys. B {\bf 581} (2000) 309.

\bibitem{dSbw}
 M.K. Parikh and S.N. Solodukhin,
    {\it De Sitter brane gravity: From close-up to panorama},
    Phys. Lett. B {\bf 503} (2001) 384;
 A. Wang,
    {\it Thick de Sitter 3-Branes, Dynamic Black Holes and Localization of Gravity},
     Phys. Rev. D {\bf 66} (2002) 024024;
     S. Nojiri and S.D. Odintsov,
     {\it Quantum cosmology, inflationary brane world creation and dS/CFT correspondence},
     JHEP {\bf 0112} (2001) 033;
     S. Nojiri, S.D. Odintsov and S. Ogushi,
     {\it Graviton correlator and metric perturbations in de Sitter brane world},
     Phys. Rev. D {\bf 66} (2002) 023522;
     I.H. Brevik, K. Ghoroku, S.D. Odintsov and M. Yahiro,
     {\it Localization of gravity on brane embedded in $AdS_5$ and $dS_5$},
     Phys. Rev. D {\bf 66} (2002) 064016.

\bibitem{varios}
 R. Emparan, R. Gregory and C. Santos,
    {\it Black holes on thick branes},
    Phys. Rev. D {\bf 63} (2001) 104022;
 R. Guerrero, A. Melfo and N. Pantoja,
    {\it Selfgravitating domain walls and the thin wall limit},
    Phys. Rev. D {\bf 65} (2002) 125010;
 A. Melfo, N. Pantoja and A. Skirzewski,
    {\it Thick domain wall space-time with and without reflection symmetry},
    Phys. Rev. D {\bf 67} (2003) 105003;
 K.A. Bronnikov and B.E. Meierovich,
    {\it A general thick brane supported by a scalar field},
    Grav. Cosmol. {\bf9} (2003) 313;
 O. Castillo--Felisola, A. Melfo, N. Pantoja and A. Ramirez,
    {\it Localizing gravity on exotic thick three-branes},
    Phys. Rev. D {\bf 70} (2004) 104029.
    
\bibitem{NonLocalizedFermion2}
 R. Koley and S. Kar,
    {\it A novel braneworld model with a bulk scalar field},
    Phys. Lett. B {\bf 623} (2005) 244;
    [Erratum {\it ibid}. {\bf631} (2005) 199].
 S. Pal and S. kar
    {\it De Sitter branes with a bulk scalar},
    Gen. Rel. Grav. {\bf41},1165-1179 (2009)

\bibitem{DubovskyPRD2000}
 S.L. Dubovsky, V.A. Rubakov and P.G. Tinyakov,
    {\it Brane world: disappearing massive matter},
    Phys. Rev. D {\bf 62} (2000) 105011.

\bibitem{PRD0709.3552}
 N. Barbosa-Cendejas, A. Herrera-Aguilar, M.A. Reyes Santos and C. Schubert,
    {{\it Mass gap for gravity localized on Weyl thick branes}},
    Phys. Rev.D {\bf 77} (2008) 126013.

\bibitem{ThickBrane4}
 N. Barbosa-Cendejas, A. Herrera-Aguilar, K. Kanakoglou, U. Nucamendi and I. Quiros,
    {{\it Mass hierarchy and mass gap on thick branes with Poincar\'{e} symmetry}},
    arXiv:0712.3098[hep-th].
    
\bibitem{0910.0363}
 A. Herrera-Aguilar, D. Malag\'on-Morej\'on, R.R. Mora-Luna and U. Nucamendi,
     {\it Aspects of thick brane worlds: 4D gravity localization, smoothness, and mass gap},
    Mod. Phys. Lett. A {\bf 25} (2010) 2089.


\bibitem{Liu_2010}
 Y.-X. Liu, C.-E. Fu, H. Guo, S.-W. Wei and Z.-H. Zhao,
      {\it Bulk Matters on a GRS-Inspired Braneworld},
      JCAP {\bf 1012} (2010) 031.

\bibitem{1009.1684}
 A. Herrera-Aguilar, D. Malag\'on-Morej\'on and R.R. Mora-Luna,
      {\it Localization of gravity on a thick braneworld without scalar fields},
      JHEP {\bf 1011} (2010) 015.

\bibitem{germanetal}
G. Germ\'an, A. Herrera--Aguilar, D. Malag\'on--Morej\'on, R. R.
Mora--Luna and  R. da Rocha, {\it A de Sitter tachyon thick
braneworld}, JCAP {\bf  1302} (2013) 035.

\bibitem{sen}
A. Sen, {\it Rolling tachyon}, JHEP {\bf 0204}, (2002), 048; A. Sen,
{\it Tachyon matter}, JHEP {\bf 0207}, (2002), 065; A. Sen, {\it
Field theory of tachyon matte}, Mod.Phys.Lett. A{\bf 17}, (2002),
1797-1804.

\bibitem{1401.0999}
 A. Herrera--Aguilar, A. D. Rojas and E. Santos, 
 {\it Localization of gauge fields in a tachyonic de Sitter thick braneworld},
  Phys. J. C {\bf 74} (2014); arXiv:1401.0999 [hep-th].
   
\bibitem{Corradini}
C.A. Vaquera-Araujo and O. Corradini, {\it Localization of Abelian Gauge Fields on Thick Branes},
arXiv:1406.2892 [hep-th].
   
\bibitem{Proca}
A. Escalante, C.L. Pando Lambruschini and P. Cavildo, 
{\it Hamiltonian Dynamics for Proca's theories in five dimensions with a compact dimension},
arXiv:1402.3016 [math-ph];
%
A. Escalante and A. L—pez-Villanueva,
{\it Hamiltonian dynamics of 5D Kalb-Ramond theories with a compact dimension},
arXiv:1406.0839 [hep-th].

\bibitem{FJ}
A. Escalante and M. Z\'arate, 
{\it Dirac and Faddeev-Jackiw quantization of a 5D St\"ueckelberg theory with a compact dimension},
e-Print: arXiv:1406.4430 [math-ph].

\bibitem{popeetal}
I.Y. Park, C.N. Pope, and A. Sadrzadeh, {\it AdS brane world
Kaluza-Klein reduction}, Class. Quant. Grav. {\bf 19} (2002) 6237,
arXiv: hep-th/0110238.

\bibitem{burgess}
C.P. Burgess, {\it Lectures on Cosmic Inflation and its Potential
Stringy Realizations}, Class. Quant. Grav. {\bf 24} (2007) S795,
arXiv:0708.2865 [hep-th].

\bibitem{SC}
L. McAllister and E. Silverstein, {\it String Cosmology: A Review},
Gen. Rel. Grav. {\bf 40} (2008) 565, arXiv:0710.2951 [hep-th].

\bibitem{BajcPLB2000}
 B. Bajc and G. Gabadadze,
    {\it Localization of matter and cosmological on a brane in anti de Sitter space},
    Phys. Lett. B {\bf 474} (2000) 282.

\bibitem{Liu0708}
 Y.-X. Liu, L.-D. Zhang, L.-J. Zhang and Y.-S. Duan,
    {\it {Fermions on Thick Branes in the Background of Sine-Gordon Kinks}},
    Phys. Rev. D {\bf 78}, 065025 (2008).

\bibitem{NonLocalizedFermion}
 Y. Grossman and N. Neubert,
    {\it Neutrino masses and mixings in non-factorizable geometry},
    Phys. Lett. B {\bf 474} (2000) 361.

\bibitem{LiuJCAP2009}
  Y.-X. Liu, Z.-H. Zhao, S.-W. Wei and Y.-S. Duan,
    {\it Bulk Matters on Symmetric and Asymmetric de Sitter Thick Branes},
    JCAP {\bf 02} (2009) 003.

\bibitem{Guo_jhep}
 Y.-X. Liu, H. Guo, C.-E Fu and J.-R. Ren,
    {\it Localization of Matters on Anti-de Sitter Thick Branes},
    JHEP {\bf1002} (2010) 080.

\bibitem{Ramallo}
	A.V. Ramallo, {\it Introduction to the AdS/CFT correspondence}, 
	arXiv:1310.4319 [hep-th].

\end{thebibliography}
\end{document}